\documentclass[aps,pr2,twocolumn,superscriptaddress,showpacs]{revtex4}

\usepackage{graphicx, latexsym, verbatim}
\usepackage{amssymb, amsmath}
\usepackage[ansinew]{inputenc}
\usepackage{color}

\newcommand{\ud}{\mathrm{d}}

\newcommand{\Hb}{\mathbf{H}}
\newcommand{\T}{\mathcal{T}}
\newcommand{\Kb}{\mathbf{K}}

\newcommand{\Mb}{\mathbf{M}}

\newcommand{\Sb}{\mathbf{S}}

\newcommand{\bEqa}{\begin{eqnarray}}
\newcommand{\eEqa}{\end{eqnarray}}

\begin{document}

%\preprint{0.1}

%\title{Modeling thermoelectric properties in surface disordered silicon nanowires}
\title{Surface decorated silicon nanowires: a route to high-ZT thermoelectrics}



\author{Troels Markussen}
%\email[]{tma@mic.dtu.dk}
%\thanks{}
\affiliation{Department of Micro- and Nanotechnology, Technical University of Denmark, DTU Nanotech, Building 345 East, DK-2800 Kgs. Lyngby, Denmark}

\author{Antti-Pekka Jauho}
\affiliation{Department of Micro- and Nanotechnology, Technical University of Denmark, DTU Nanotech, Building 345 East, DK-2800 Kgs. Lyngby, Denmark}
\affiliation{Department of Applied Physics, Helsinki University of Technology, P.O. Box 1100, FIN-02015 HUT, Finland}

\author{Mads Brandbyge}
\affiliation{Department of Micro- and Nanotechnology, Technical University of Denmark, DTU Nanotech, Building 345 East, DK-2800 Kgs. Lyngby, Denmark}

\date{\today}

\begin{abstract}
Based on atomistic calculations of electron and phonon transport, we
propose to use surface decorated Silicon nanowires (SiNWs) for
thermoelectric applications. Two examples of surface decorations are
studied to illustrate the underlying ideas: Nanotrees and alkyl
functionalized SiNWs. For both systems we find, (i) that the phonon
conductance is significantly reduced compared to the electronic
conductance leading to high thermoelectric figure of merit, $ZT$,
and (ii) for ultra-thin wires surface decoration leads to
significantly better performance than surface disorder.
\end{abstract}

% 72.10.Fk Scattering by point defects, dislocations, surfaces, and other imperfections (including Kondo effect)
% 72.15.Lh Relaxation times and mean free paths
% 73.63.-b Electronic transport in nanoscale materials and structures (see also 73.23.-b Electronic transport in mesoscopic systems)
% (73.21.Hb Quantum wires)

% 72.10.Fk Scattering by point defects, dislocations, surfaces, and other imperfections (including Kondo effect)
% 72.15.Lh Relaxation times and mean free paths
% 73.63.-b Electronic transport in nanoscale materials and structures (see also 73.23.-b Electronic transport in mesoscopic systems)
% 72.15.Jf 	Thermoelectric and thermomagnetic effects 
% (73.21.Hb Quantum wires)
% 63.22.-m Phonons or vibrational states in low-dimensional structures and nanoscale materials  
% 63.22.Gh Nanotubes and nanowires  
% 66.70.-f Nonelectronic thermal conduction and heat-pulse propagation in solids; thermal waves  

\pacs{63.22.Gh, 72.15.Jf, 73.63.-b, 66.70.-f}

%\keywords{}

\maketitle
Recent ground-breaking experiments indicate that rough silicon
nanowires (SiNWs) can be efficient thermoelectric materials although
bulk silicon is not \cite{HochbaumNature2008,BoukaiNature2008}: they
conduct charge well but have a low heat conductivity. 
A measure for the performance is given by the figure of merit $ZT=G_e S^2
T/\kappa$, where $G_e$, $S$, and $T$, are the electrical
conductance, Seebeck coefficient, and (average) temperature,
respectively. The heat conduction has both electronic and phononic
contributions, $\kappa=\kappa_e+\kappa_{ph}$. Materials with
$ZT\sim1$ are regarded as good thermoelectrics, but $ZT>3$ is
required to compete with conventional refrigerators or
generators~\cite{MajumdarScience2004}. Recent theoretical works predict $ZT>3$ in ultra-thin SiNWs~\cite{VoNanoLett2008,KnezevicIEEE08,MarkussenPRB2009}. 
The high performance SiNWs in Ref. \cite{HochbaumNature2008} were
deliberately produced with a very rough surface, and the high $ZT$
is attributed to increased phonon-surface scattering which decreases
the phonon heat conductivity, while the electrons are less affected
by the surface roughness. The extraordinary low thermal conductivity measured in rough SiNWs is supported be recent calculations~\cite{MartinPRL2009,DonadioGalliPRL2009}.
Surface disorder will, however, begin to
affect the electronic conductance significantly in very thin wires
\cite{PerssonNanoLett2008} and thereby reduce
$ZT$~\cite{KnezevicIEEE08,MarkussenPRB2009}.

%%\section{Introduction}

\begin{figure}[htb!]
    \begin{center}
    \begin{minipage}[c]{\columnwidth}
        \includegraphics[width=0.75\columnwidth, angle=0]{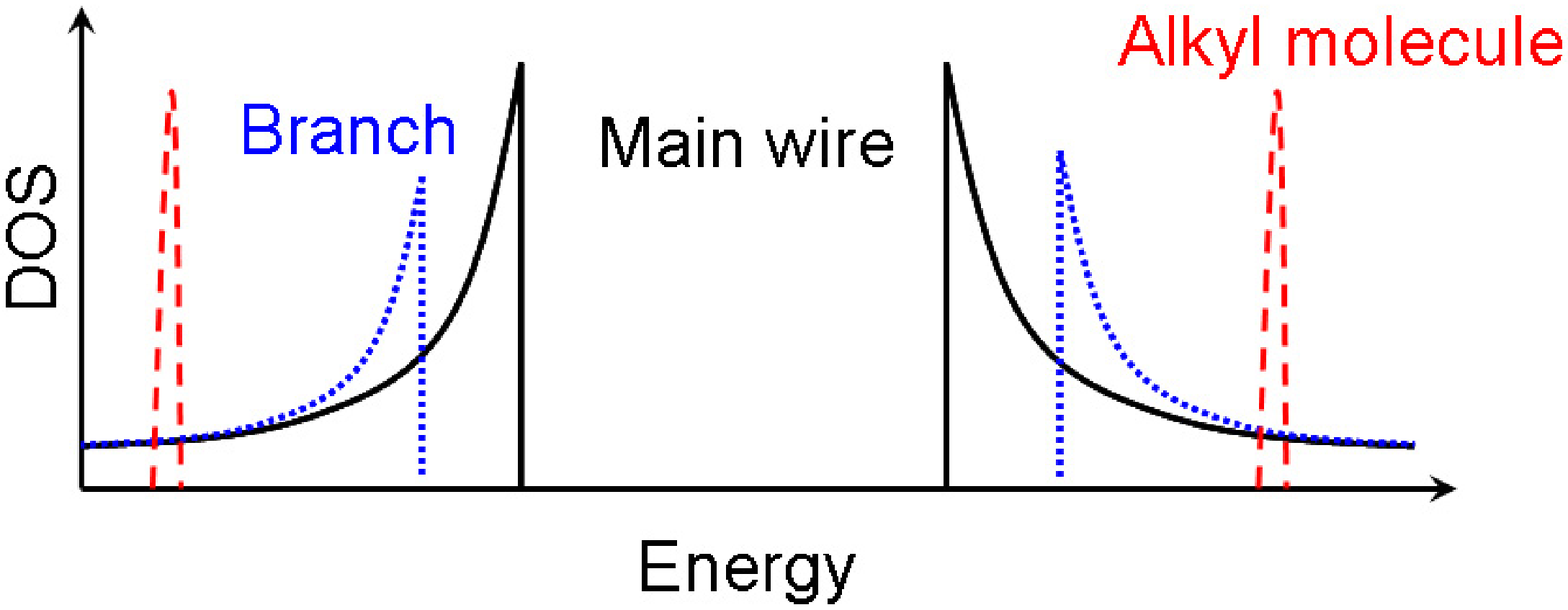}
        \vspace{5mm}
    \end{minipage}
    \begin{minipage}[c]{\columnwidth}
        \includegraphics[width=0.75\columnwidth, angle=0]{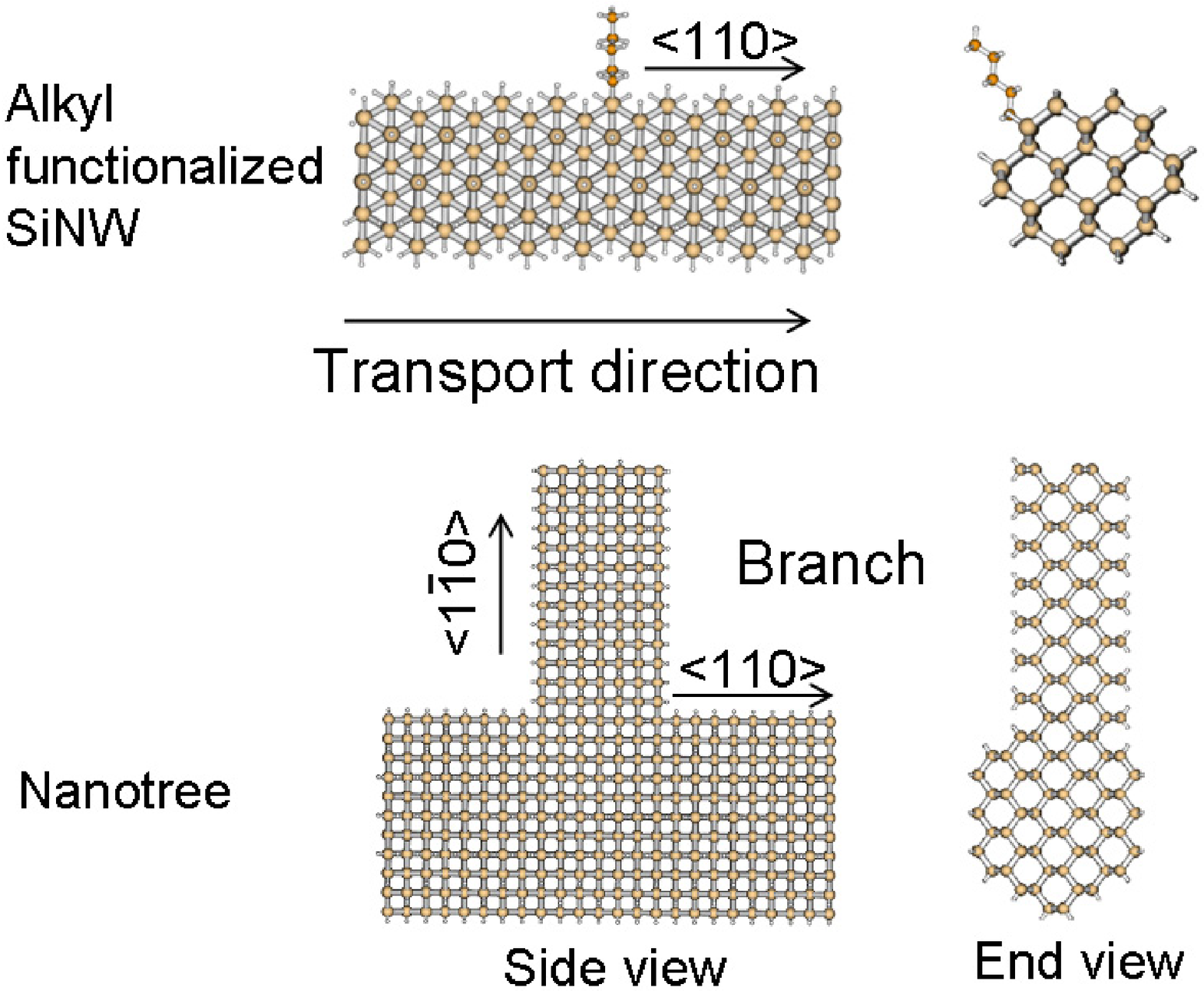}
        \end{minipage}
    \end{center}
    \caption{Sketch of electronic density of states (top) in the alkyl functionalized
   SiNW (middle) and in the nanotree (bottom). The trunk in the nanotree is
   broader than the branch and thus has a smaller bandgap.
   Also the alkyl molecular states are located deep inside the bands.
   For both structures electrons and holes close to the band edges are therefore only weakly scattered
   while phonons are strongly scattered.}
    \label{nanoTreePrinciple}
\end{figure}

For defect-free wires, the room temperature phononic conductance
scales with the cross-sectional area
\cite{MarkussenPristinePhononPaper}. Contrary, for ultra-thin wires
with diameters $D\lesssim 5\,$nm, the electronic conductance is not
proportional to the area: it is quantized and given by the number of
valence/conduction band states at the band edges. Decreasing the
diameter to this range thus decreases $\kappa_{ph}$ while keeping $G_e$ almost constant. An optimally designed thermoelectric
material would scatter phonons but leave the electronic conductance
unaffected, even for the smallest wires. Instead of introducing
surface disorder, which affects the electronic conductance in the
thin wires, it is interesting to speculate whether other surface
designs could lead to improved thermoelectric performance. Lee {\it et al.}~\cite{LeeNanoLett2008} proposed nanoporous Si as an efficient thermoelectric matrial. Blase and
Fern\'andez-Serra~\cite{BlasePRL2008} recently showed that covalent
functionalization of SiNW surfaces with alkyl molecules leaves the
electronic conductance unchanged, because the molecular states are
well separated in energy from the nanowire bandedges, see Fig.
\ref{nanoTreePrinciple}. If the alkyl molecules scatter the phonons,
such functionalized SiNWs would be candidates for thermoelectric
applications. Experimental alkyl functionalization of SiNWs was reported in Ref. \cite{Lewis2006}.

Another possible surface design are branched SiNWs, so-called
nanotrees. Nanotrees have been synthesized in III-VI
semiconductors~\cite{DickJCrystGrowth2004,DickNMat2004} and in
silicon~\cite{FonsecaAPL2005,Doerk2008}. The stability and
electronic structure of silicon nanotrees have recently been
addressed theoretically~\cite{Menon2007,AvramovNanoLett2007}. The
thinner branches will have a larger band gap than the main wire
("trunk"), see Fig. \ref{nanoTreePrinciple}. Close to the band
edges, the electronic scattering is therefore weak.

The presence of an alkyl molecule or a nanowire branch leads both to
a reduction, $\Delta\kappa$, of the thermal conductance and a
reduction, $\Delta G$, of the electronic conductance. In this letter
we show that at room temperature (RT) the ratio $\Delta\kappa/\Delta
G>50$ for the alkyl functionalized SiNWs, and $\Delta\kappa/\Delta
G>20$ for a nanotree. By engineering the SiNW surfaces it is thus
possible to reduce the phonon conductance while keeping the
electronic conductance almost unaffected. We thus propose such
surface decorated SiNWs as promising candidates for nanoscale
thermoelectric applications.

%********************************************************************************************
%********************************************************************************************

\textit{Systems.} We consider two specific systems shown in Fig.
\ref{nanoTreePrinciple}. The first is an alkyl functionalized SiNW
with a wire diameter of 12~\AA, and with the wire oriented along the
$\langle110\rangle$ direction. The alkyl (C$_n$H$_{2n+1}$) is
attached to the H-passivated nanowire replacing a H atom. The second
system is a nanotree, where a small diameter (12 \AA) branch is
attached to a larger diameter (20 \AA) trunk. The trunk is oriented
in the $\langle110\rangle$ direction while the branch is oriented
along the $\langle1\bar{1}0\rangle$ direction, and is thus
perpendicular to the trunk. The length of the branch, $L_B$, is
varied.

\textit{Methods.} The electronic Hamiltonian, $\Hb$, and overlap
matrix, $\Sb$, of the alkyl functionalized SiNWs are obtained from
local orbital DFT calculations~\cite{siesta-ref,dft-details}. The
calculations are performed on super-cells containing 7 wire unit
cells with the alkyl molecule bound to the middlemost unit cell, as
shown in Fig. \ref{nanoTreePrinciple}.

%The SiNW is again oriented along the $\langle110\rangle$ direction
For the nanotrees, we use a tight-binding (TB) model since these
systems contain $>1100$ atoms, too many for our DFT implementation.
The electronic TB Hamiltonian describing the nanotree is calculated
using a 10 band $sp^3d^5s^*$ nearest-neighbor orthogonal TB
parametrization~\cite{Boykin2004,Lake2005}. We recently applied the same TB model to study thermoelectric properties of surface disordered SiNWs~\cite{MarkussenPRB2009}.
%The same TB parameters were recently applied to study SiNW band structures~\cite{NiquetPRB2006}, surface roughness \cite{PerssonNanoLett2008,LuisierAPL2007}, and thermoelectric properties of surface disordered SiNWs~\cite{MarkussenPRB2009}.

The phononic system, characterized by the force constant matrix,
$\Kb$, is described using the Tersoff empirical potential (TEP)
model~\cite{Tersoff1988,Tersoff1989} for both the nanotree and the
functionalized SiNW. For pristine wires, we have recently shown that
the TEP model agrees well with more elaborate DFT
calculations~\cite{MarkussenPristinePhononPaper}. We limit our
description to the harmonic approximation, thus neglecting
phonon-phonon scattering. The harmonic approximation is always valid
at low temperatures. In bulk Si, the room temperature anharmonic
phonon-phonon relaxation length at the highest frequencies is
$\lambda_a(\omega_{max})\sim 20\,$nm and increases as
$\lambda_a\propto\omega^{-2}$ at lower frequencies
\cite{MingoYangPRB2003}. Experimental studies of silicon
films~\cite{JuAPL1999} showed that the effective mean free path of
the dominant phonons at room temperature is $\sim$300 nm. For
relatively short wires with lengths $L\lesssim 100\,$nm the
anharmonic effects thus seem to be of limited importance, and the
harmonic approximation is expected to be good.

We calculate the electronic conductance from the electronic
transmission function, $\mathcal{T}_e(\varepsilon)$. This is
obtained from the $\Hb,\Sb$-matrices following the standard
non-equilibrium Green's function (NEGF)/Landauer setup, where the
scattering region (i.e. the regions shown in Fig.
\ref{nanoTreePrinciple}) is coupled to semi-infinite, perfect
wires~\cite{Haug08}. The electronic quantities in the $ZT$ formula
can be written
as~\cite{SivanImry1986,EsfarjaniPRB2006,LundeFlensberg2005}
$G_e=e^2L_0$, $S=L_1(\mu)/[eT\,L_0(\mu)]$ and
$\kappa_e=[L_2(\mu)-(L_1(\mu))^2/L_0(\mu)]/T$ where $L_m(\mu)$ is
given by
\begin{equation}
L_m(\mu) = \frac{2}{h}\int_{-\infty}^\infty \ud \varepsilon\,\mathcal{T}_e(\varepsilon)(\varepsilon-\mu)^m\left(-\frac{\partial f(\varepsilon,\mu)}{\partial \varepsilon}\right) \label{Lm}.
\end{equation}
Here $f(\varepsilon,\mu) = 1/\left(\exp\left[(\varepsilon-\mu)/k_BT\right]+1\right)$
 is the Fermi-Dirac distribution function at the chemical potential $\mu$.

The phonon transmission function, $\mathcal{T}_{ph}(\omega)$, at
frequency $\omega$ is calculated in a similar way as the electronic
transmission with the substitutions $\Hb\rightarrow\Kb$ and
$\varepsilon\Sb\rightarrow \omega^2\Mb$, where $\Mb$ is a diagonal
matrix with the atomic masses~\cite{YamamotoPRL2006,MingoPRB2006,WangPRB2006}. The phonon thermal conductance is
\begin{equation}
\kappa_{ph}(T) = \frac{\hbar^2}{2\pi k_B T^2}
\int_{0}^\infty\ud\omega\,\omega^2\,\mathcal{T}_{ph}(\omega)\,
\frac{e^{\hbar\omega/k_BT}}{(e^{\hbar\omega/k_BT}-1)^2}. \label{ThermalConductance}
\end{equation}

%********************************************************************************************
%********************************************************************************************

\begin{figure}[htb!]
    \begin{center}
        \includegraphics[width=0.9\columnwidth, angle=0]{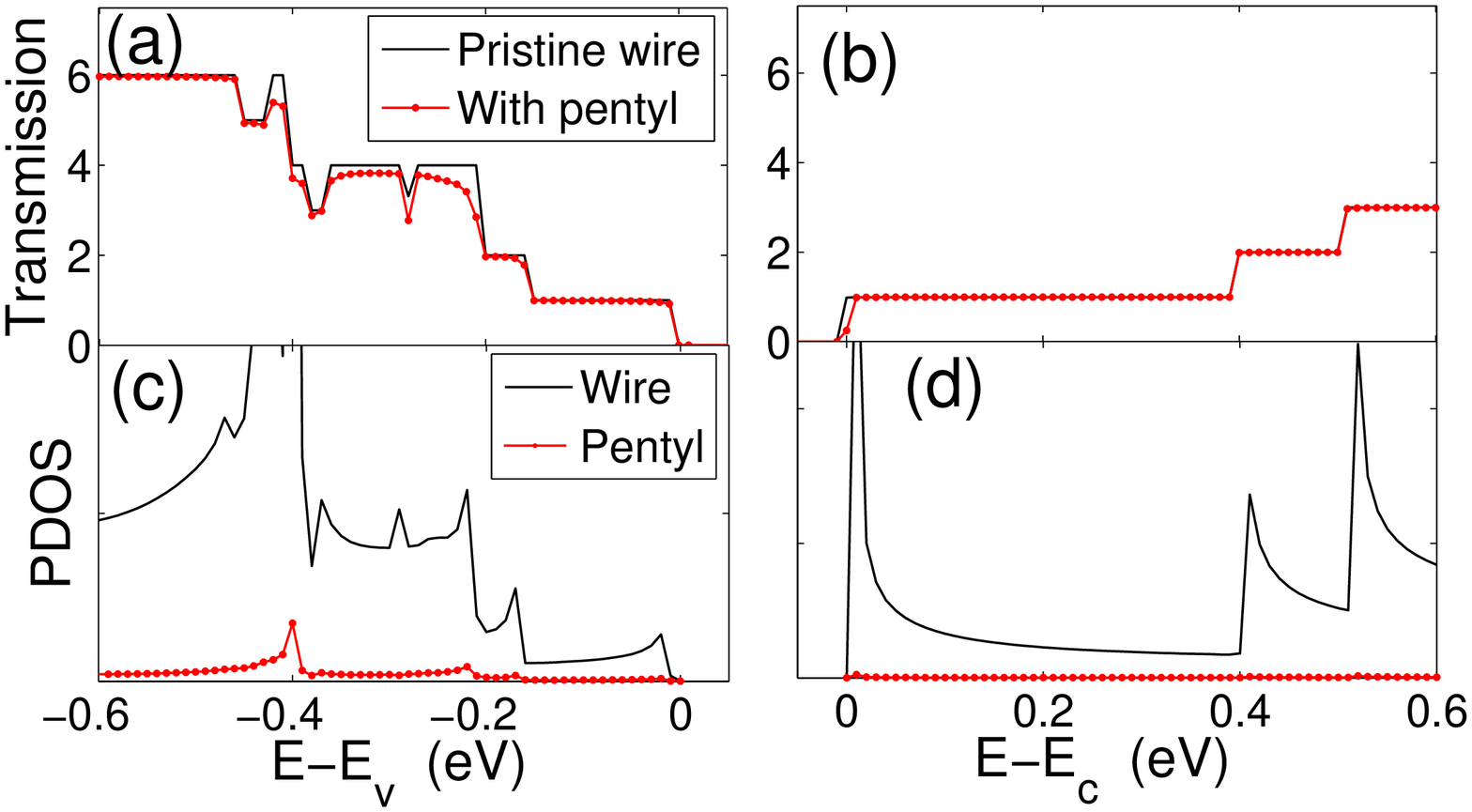}   %L:/Matlab/yans/functionalized/results
        \includegraphics[width=0.9\columnwidth, angle=0]{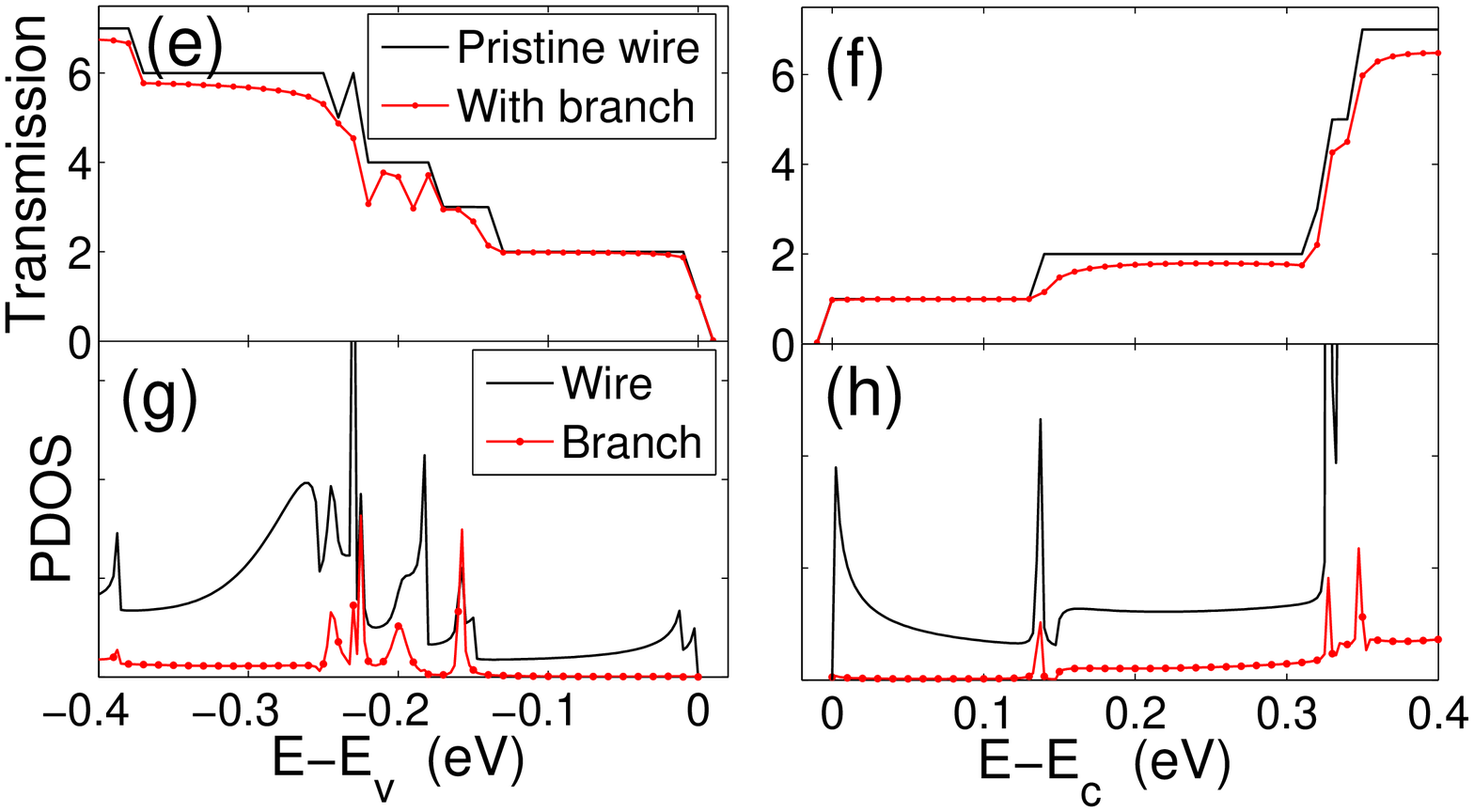}             % L:/Matlab/yans/nanoTree/results/figures/
    \end{center}
    \caption{Hole and electron transmissions for pentyl functionalized,
   $D=12\,$\AA~SiNW (a)-(b) and for a nanotree with $D=20\,$\AA~trunk and $D=12\,$\AA~branch (e)-(f).
   Panels (c)-(d) and (g)-(h) show the PDOS on the wire and on the pentyl and branch, respectively. The energy scales are relative to the valence band edge, $E_v$  for the holes (left column) and relative to the conduction band edge, $E_c$ for electrons (right column).The DFT band gap of the $D=12\,$\AA~wire is 1.65 eV, while the TB band gap of the $D=20\,$\AA~wire is 1.77 eV.}
    \label{elecTrans}
\end{figure}

\textit{Charge transport.} Figures \ref{elecTrans} (a) and (b) show
the calculated hole and electron transmissions for a pentyl
(C$_5$H$_{11}$) functionalized SiNW. Notably, the transmission is
nearly perfect close to the bandedges. The average reduction of the
transmission in the first hole and electron conductance plateaus are
2\% and 0.2\%, respectively, in agreement with the findings of
Ref.~\cite{BlasePRL2008}. Figures 2 (c) and (d) show the projected density of states (PDOS) on the wire and on the pentyl. The high transmission regions in panel (a) and (b) are seen to correspond with regions of vanishing PDOS on the pentyl molecule. Likewise, at energies in the valence band where scattering is observed, there is a relatively large PDOS at the pentyl.

Figures 2 (e) and (f) show the transmission through the nanotree.
The branch length is $15.4\,$\AA. Again,  the transmission close to
the band edges is nearly perfect, with a reduction of 2\% and 0.9\%
for holes and electrons, respectively. Figures 2 (g) and (h) show the PDOS on the main wire and on the branch. We again observe a correspondence between perfect transmission and low PDOS on the branch.

The almost perfect transmissions close to the band edges
can be qualitatively understood from the schematic drawing in Fig.
\ref{nanoTreePrinciple} (top). The HOMO and LUMO level of the pentyl
are located deep inside the bands~\cite{BlasePRL2008} and the
molecular states are thus not accessible for electrons or holes
close to the band edges. For the nanotree, the branch has a smaller
diameter and thus a larger bandgap. Electrons or holes in the trunk,
with energies close to the band edges, are not energetically allowed
in the branch and therefore do not 'see' the branch. In addition to
the energy considerations, the spatial distribution of the Bloch
state also plays a role: the first valence and conduction band Bloch
states of the main wire have more weight in the center of the wire
than at the edge~\cite{MarkussenPRB2009}.

\begin{figure}[htb!]
    \begin{center}
        \includegraphics[width=0.9\columnwidth, angle=0]{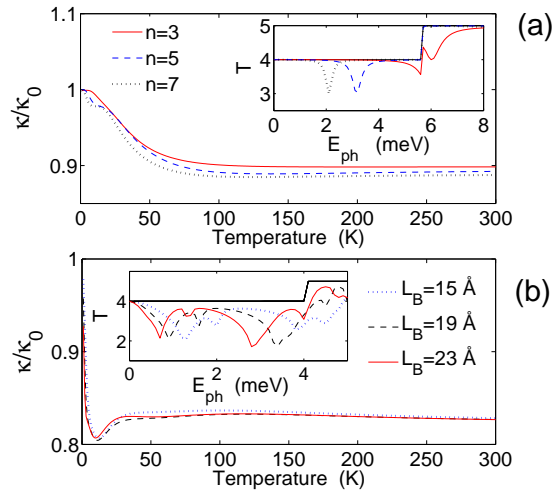}
    \end{center}
    \caption{Thermal conductance ratio $\kappa/\kappa_0$ for (a) pentyl
functionalized SiNWs with different diameters, and
(b) for nanotrees with different branch lengths, $L_B$.
The nanotree trunks are again $D=20\,$\AA~ with $12\,$\AA~diameter branches. The insets show the phonon transmission function at low energies. Fano-like resonant scattering is observed in both systems.}
    \label{kk0Fig}
\end{figure}

\textit{Thermal transport.} Figure \ref{kk0Fig} shows the
temperature dependence of the thermal conductance ratios
$\kappa/\kappa_0$, where $\kappa_0$ is the pristine wire thermal
conductance, which in the low energy limit equals the universal thermal conductance quantum, $\kappa_Q(T)=4(\pi^2\,k_B^2T/3h)$~\cite{SchwabNature2000}.
Panel (a) shows the ratios for wires with alkyles, C$_n$H$_{2n+1}$,
with different lengths, $n=3,5,7$. The thermal conductance at RT is
reduced by $\sim10$\%, and the overall behavior does not depend on
the alkyl length. The inset shows the phonon transmission at low
phonon energies. Note the resonant dips in the transmission, where
exactly one channel is closed yielding a transmission of three.
These dips are associated with an increased local phonon density of
states at the alkyl molecule at the resonant energies, corresponding
to a localized vibrational mode. Such Fano-like resonant scattering
is well-known from electron transport~\cite{NockelStone}. A phonon eigenchannel analysis~\cite{PaulssonPRB2007} shows that the transmission dips are due to a complete blocking of the rotational mode in the wire. The corresponding localized alkyl phonon mode is a vibration in the plane perpendicular to the wire axis.

Panel (b) in Fig. \ref{kk0Fig} shows the thermal conductance ratio
for nanotrees with different branch length, $L_B$. There is only a
weak dependence on $L_B$ at low temperatures, and at RT the four
curves basically coincide showing a thermal conductance reduction of
17\% of the nanotree compared to the pristine wire. Again we observe
resonant transmission dips for the nanotrees. Two channels - the rotational and one flexural mode - close completely at the resonance due to two quasi-localized vibrational modes in the branch.  
These phonon backscattering resonances are
responsible for the dip in the $\kappa/\kappa_0$ ratio around
$T=10\,$K. Notice that all the conductance ratios approach unity in
the low temperature limit. This is because the four acoustic modes
transmit perfectly in the limit $\omega\rightarrow
0$~\cite{SchwabNature2000}.

\begin{figure}[htb!]
    \begin{center}
        \includegraphics[width=0.75\columnwidth, angle=0]{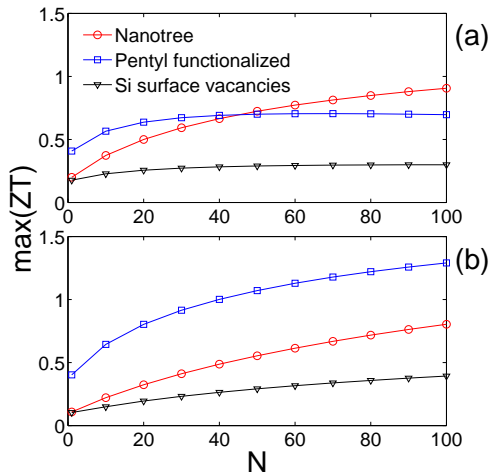}
        %\put(-60,175){$p$-type}
        %\put(-160,85){$n$-type}
    \end{center}
    \caption{Thermoelectric figure of merit, $ZT$, for $p$-type (a) and $n$-type (b) wires.
    $N$ is the number of pentyl molecules (squares), nanotree branches (circles) and silicon surface vacancies (triangles) in the wire.}
    \label{ZTfig}
\end{figure}

We may vary the thermoelectric figure of merit, $ZT$, by varying the chemical potential.
Typically $ZT$ displays a maximum for $\mu$ close to the band
edge~\cite{MarkussenPRB2009,VoNanoLett2008}. Figure \ref{ZTfig} shows the maximum $ZT$
values for the pentyl functionalized SiNW (squares), the nanotree (circles), and surface
disordered SiNWs (triangles), where disorder is modeled by introducing surface silicon
vacancies. The diameter of the surface disordered wire is $D=20\,$\AA~and it is oriented
along the $\langle110\rangle$ direction. The calculational details are given in
Ref.~\cite{MarkussenPRB2009}. The curves show $ZT$ as a function of the number ($N$) of
pentyl molecules/nanotree branches/silicon vacancies. In calculating $ZT$ vs $N$, we
have assumed that the transmission, $\mathcal{T}_N$, through a longer wire with e.g. $N$
pentyl molecules randomly covering the surface can be obtained from the single-pentyl
transmission, $\T_1$ as $\T_N^{-1} = \T_0^{-1} + N(\T_1^{-1}-\T_0^{-1})$, where $\T_0$
is the pristine wire transmission. The term in parenthesis corresponds to a scattering
resistance of a single pentyl molecules. This averaging method has recently been
validated in the quasi-ballistic and diffusive regimes for both electron and phonon
transport~\cite{MarkussenPRB2009,SavicMingoPRB2008,MarkussenPRL2007}.

Figure \ref{ZTfig} shows that increasing the number of scattering
centers, i.e. the number of pentyl molecules or nanotree branches
increases the $ZT$ for both hole transport (a) and electron
transport (b). In the case of holes in the pentyl functionalized
SiNWs, the $ZT$ reaches an almost constant level of $ZT=0.7$ at
$N=40$, but in all other cases, $ZT$ increases throughout the range.
Increasing the density of molecules/nanotree branches or increasing
the length of the wire will thus increase the thermoelectric
performance. The reason is that the electrons (holes) are less
affected by the surface modifications than the phonons, as also seen
in Figs. \ref{elecTrans} and \ref{kk0Fig}. The surface disordered
wires (triangles) show an increasing $ZT$ vs $N$ but at values
significantly lower than the two other surface modified wires.

\textit{Discussion.} A number of idealizations have been made in
our calculations, and we next assess their significance.  The
structures we have considered represent plausible choices, dictated
by computational limitations, but do not necessarily match
quantitatively real structures.  Thus, for example, a surface
decorated SiNW will also be rough, and one should consider the
combined effect of all scattering mechanisms. We have not carried
out optimizations neither with respect to the attached molecules nor
with respect to the geometry of the nanotrees. Electron-phonon and
phonon-phonon scattering  will affect both the electronic and
thermal conductances and the obtained $ZT$
values~\cite{KnezevicIEEE08}.  We do not expect to reach
quantitative agreement with experiment but believe to have
identified important trends: In  SiNW based thermoelectrics, surface
decorations in terms of added molecules or nanowire branches seem to
be a better approach than surface disorder in the ultra-thin limit.

We thank the Danish Center for Scientific Computing (DCSC) and
Direkt\o r Henriksens Fond for providing computer resources. TM
acknowledges the Denmark-America foundation for financial support.
APJ is grateful to the FiDiPro program of the Finnish Academy.

%\bibliography{bib1}

\begin{thebibliography}{39}
\bibitem[{\citenamefont{aa}(2009)}]{dft-details}
\bibinfo{note}{We   use a single-$\zeta$ polarized basis set with an energy cutoff of 200 Ry,
  norm-conserving pseudopotentials and the generalized-gradient approximation
  for the exchange-correlation functional.}
\end{thebibliography}

\end{document}